\documentclass[12pt]{iopart}
\usepackage{graphicx,iopams}
\begin{document}
\bibliographystyle{unsrt}  
\title[On the kink confinement in $\mathrm{Co}\mathrm{Nb}_2\mathrm{O}_6$  ]
{On the weak confinement of kinks in the one-dimensional quantum ferromagnet $\mathrm{Co}\mathrm{Nb}_2\mathrm{O}_6$  }
\author{S.~B. Rutkevich}
\address{Institute of Solid State and Semiconductor Physics,  SSPA  
"Scientific-Practical Materials Research Centre, 
NAS of Belarus", P.~Brovka St. 17, 220072 Minsk, Belarus}
\begin{abstract}
In a recent paper Coldea \etal (2010 Science {\bf 327}  177) report observation
of the weak confinement of kinks in the Ising spin chain ferromagnet $\mathrm{Co}\mathrm{Nb}_2\mathrm{O}_6$
at low temperatures. To interpret the entire spectra of magnetic excitations measured via neutron scattering,
they introduce a phenomenological model, which takes into account only the two-kink configurations of the spin chain.
We present the exact solution of this model. The explicit expressions  for the two-kink bound-state energy spectra
and for the relative intensities of  neutron scattering on these magnetic modes are obtained in terms of the Bessel 
function. 	
\end{abstract}
\ead{rut@ifttp.bas-net.by}
\section{Introduction}
Very recently Coldea \etal \cite{Coldea10} reported the impressive results of 
inelastic neutron scattering experiments on the quasi-1D ferromagnetic $\mathrm{Co}\mathrm{Nb}_2\mathrm{O}_6$
(cobalt niobate)  single crystal. The essential physics of this material at low temperatures can be described 
in terms of the quantum Ising spin chain model, which is paradigmatic for the theory of the quantum phase 
transitions \cite{Sach99}. In presence of the magnetic field $h_\perp$ transverse to the easy magnetization axis, 
the ground state state of the model can be either ferromagnetic or paramagnetic depending on the strength of the
external magnetic field $h_\perp$.
The transition between the two phases occurs (at zero temperature) at the critical value of the transverse field
$h_\perp=h_c$. This critical point belongs to the 2D Ising universality class.  
 
The  spectra observed by Coldea \etal \cite{Coldea10} display certain very subtle features
providing experimental confirmation of two long-standing theoretical predictions \cite{McCoy78,ZamH},
which  relate to the 
Ising model. Directly at the critical transverse field $h_\perp=h_c=5.5 T$, the
ratio of two lightest quasiparticles approaches to the 'golden ratio' indicating the hidden $E_8$ symmetry in the 
critical Ising model in the longitudinal magnetic field, 
as it was predicted by A.B. Zamolodchikov \cite{ZamH} nearly two decades ago. 
On the other hand, at zero magnetic field, the observed energies of five lowest magnetic excitations 
were proportional to the absolute values $z_n$ of zeroes 
of the Airy function, $\mathrm{Ai}(-z_n)=0$, in agreement with the 
theory of the kink confinement originating in 1978 from the work of McCoy and Wu \cite{McCoy78}. On the
resent developments in this field
 see \cite{Del96,DelMus98,FonZam2003,FZ06,DG08,LTD09,Rut09},
further references can be found in the monograph \cite{Mussardo10}. 

Confinement of topological excitations typically takes place in two dimensions (one spacial and one time dimension), if 
the discrete vacuum degeneracy is explicitly broken by a small
interaction term.  In the simplest heuristic approach,
two confined kinks in the ferromagnetic Ising chain are treated as two quantum particles moving in a line
$-\infty<x<\infty$
and attracting one another with a linear potential $\lambda |x|$. 
The latter can be induced in the quasi-1D Ising ferromagnet
either by a weak external longitudinal  magnetic field, or by the weak coupling between the magnetic chains
in the 3D magnetically ordered phase \cite{Coldea10,Tsv03,Tsv04}. In this approach, 
the relative motion of two kinks is described by
the Schr{\"o}dinger equation
\begin{equation} \label{Sch}
-\frac{1}{m}\frac{d^2}{d x^2} \psi_n(x)+ \lambda |x| \psi_n(x) =\delta E_n\,  \psi(x),
\end{equation}
with a skew-symmetric{\begin{footnote} {Equation (\ref{Sch}) 
with a {\it symmetric} wave function $\psi(x)=\psi(-x)$ can describe the two-kink bound states
 in the $3$-state
Potts field theory \cite{RutP09,LTD09}. 
} \end{footnote}} wave function $\psi(x)=-\psi(-x)$. It  immediately leads to the energy
levels of the kink bound-states \cite{McCoy78}
\begin{equation} \label{spS}
\delta E_n = z_n \lambda^{2/3} m^{-1/3}, \quad n=1,2\ldots
\end{equation}

The simple theory of confinement based on (\ref{spS}) implies the quadratic 
dispersion law $p^2/(2m)$ for a free kink, and ignores discreteness of the spin chain.
Though these approximations are reasonable for small enough momenta 
 of the composite two-kink bound states, 
a more systematic approach is required to describe their spectra in the whole Brillouin zone.

The effect of the lattice discreteness on the kink confinement in the non-critical Ising spin chain  
has been studied in ref. \cite{Rut08} in the Bethe-Salpeter equation approach \cite{FonZam2003,FZ06}.
To interpret the full experimental spectra in $\mathrm{Co}\mathrm{Nb}_2\mathrm{O}_6$, 
Coldea and his colleagues
proposed a different phenomenological model, which takes into 
account only the two-kink (i.e. one-domain) configurations of the spin chain. The 
Hamiltonian of this model is given in \cite{suppC10}  by equation (S1),
which we reproduce in equation (\ref{H}) below. The neutron scattering spectra were compared by authors of  
ref. \cite{Coldea10} with an approximate  perturbative solution  \cite{pcC10} of their phenomenological model.
It is interesting,  that this model admits an exact solution, which we describe in the present paper. 
Our main results are equations (\ref{sp}), (\ref{Bes}), and (\ref{ikinh}), which express 
the energy spectra and relative neutron-scattering intensities of 
the magnetic excitation modes for model (\ref{H}) in terms of the Bessel function.

The rest of the paper is organized as follows. In Section \ref{Sec2} the phenomenological model
introduced in \cite{suppC10} is described and its exact energy spectrum is obtained. These exact spectra
are analyzed in several asymptotical regimes in the  small magnetic field limit in Section \ref{Sec3}. Section \ref{Sec4}
contains calculation of the dynamical correlation function, which is proportional to the 
neutron-scattering intensities. The details of calculations are described in two Appendices. Concluding remarks
are presented in Section \ref{Conc}. 
\section{The two-kink model and its energy spectrum \label{Sec2}}
In this Section we study the eigenvalue problem for the model Hamiltonian defined by equation (S1) of \cite{suppC10}:
\begin{eqnarray}\nonumber
&& H  |j,l\rangle = J|j,l\rangle-\alpha[|j,l+1\rangle+(|j,l-1\rangle +
|j+1,l-1\rangle)(1- \delta_{l,1} )\\
&&
+|j-1,l+1\rangle]
+h_z l|j,l\rangle-\beta\delta_{l,1}(|j-1,1\rangle+|j+1,1\rangle)+\beta'|j,1\rangle \delta_{l,1}. \label{H}
\end{eqnarray}
Here $|j,l\rangle$ denotes the two-kink state of the ferromagnetic spin-$1/2$ chain,
\[
|j,l\rangle=|\ldots\uparrow\uparrow\uparrow\stackrel{j}{\downarrow}
\downarrow\ldots\downarrow\stackrel{j+l}{\uparrow}\uparrow\uparrow\ldots\rangle,
\] 
the indices $j$ and $l$ give the starting position and the length of the down-spin cluster,
$j=0,\pm 1,\pm 2,  \ldots$, and $l=1,2,\ldots$ Parameter $J$ characterizes the energy needed to create two kinks. 
The terms proportional to $\alpha$ describe the nearest neighbour hoppings of kinks along the chain. The long-range attraction between the kinks
is represented in  (\ref{H}) by the term $h_z l$, where $h_z$ is the effective longitudinal magnetic field. 
The short-range interaction $\beta$- and $\beta'$-terms were introduced in \cite{suppC10} to 
describe the experimentally observed 'kinetic mode' - the well localized   bound-state 
mode near the Brillouin zone boundary \cite{Coldea10}. Note, that $\beta$- and $\beta'$-terms has no analogue in the 
standard Ising spin chain Hamiltonian. 

In the momentum basis
\begin{equation}  \label{Pbas}
|P,l\rangle =\sum_{j=-\infty}^\infty\exp(\rmi Pj)|j,l\rangle,
\end{equation}
the Hamiltonian is diagonal in the momentum variable $P$
and  acts on the basis state  as follows
\begin{eqnarray}\nonumber
&& H  |P,l\rangle = -\alpha[ (1+e^{\rmi P}) |P,l+1\rangle +(1- \delta_{l,1})(1+e^{-\rmi P})|P,l-1\rangle]+\\
&&[J +h_z l +(\beta'-2 \beta \cos P)\delta_{l,1}]|P,l\rangle.
 \label{HP0}
\end{eqnarray}

The eigenvalue problem 
\begin{equation}
H |\Phi(P)\rangle  =    E(P) |\Phi(P)\rangle  \label{ei}
\end{equation}
takes in basis (\ref{Pbas}) the explicit form
\begin{eqnarray}\nonumber
&&[J +h_z l +(\beta'-2 \beta \cos P)\delta_{l,1}-E(P)] \psi(l,P) -\\
&&  -2\alpha\cos(P/2)[ \psi(l+1,P)+(1- \delta_{l,1})\psi(l-1,P)]=0,
 \label{HP}
\end{eqnarray}
where $l=1,2,\ldots$, the momentum $P$ has the Brillouin zone $-\pi<P<\pi$, and 
\[
2 \pi \,\delta(P'-P)\,\psi(l,P)=\exp(-\rmi Pl/2)\langle P',l|\Phi(P)\rangle.
\]
Problem (\ref{ei}) can be easily solved in the case of zero magnetic field. 
The energy spectrum at $h_z=0$ has the continuous part
\begin{equation} \label{cont}
E(k,P)\equiv\varepsilon(k,P)=J- 4\alpha \cos(P/2)\cos k,
\end{equation}
and one localized bound state, the 'kinetic mode',
\begin{equation}\label{kin0}
E_{kin}(P)=J+(\beta'-2\beta \cos P)+\frac{4\alpha^2 \cos^2(P/2)}{ \beta'-2\beta \cos P }.
\end{equation}
Note, that the wave functions corresponding to the continuous spectrum 
and to the kinetic bound state mode read, respectively,  as
\begin{eqnarray} 
\psi(l,P;k)=A\,\sin[k(l-1)+\gamma], \label{conti}  \\
\label{loc}
\psi_{kin}(l,P)= B\,z ^l,
\end{eqnarray}
where $A$ and $B$ are the normalizing constants, and 
\begin{eqnarray} \nonumber
z=\frac{2\alpha \cos(P/2)}{2 \beta \cos P-\beta'}\,,\\
\cot\gamma=\cot k+ \frac{\beta'-2\beta\cos P}{2\alpha\sin k\,\cos(P/2)} \label{gamma}
\end{eqnarray}

Requirement $|z|<1$ implies, that the  kinetic mode (\ref{loc}) exists only in the region near
 the Brillouin zone boundary, at 
\begin{equation}
|\cos (P/2)|<\frac{-\alpha+(\alpha^2+8\beta^2+4\beta\beta')^{1/2}}{2\beta}.
\end{equation}
\begin{figure}[htb]
\centering
\includegraphics[width=\linewidth, angle=00]{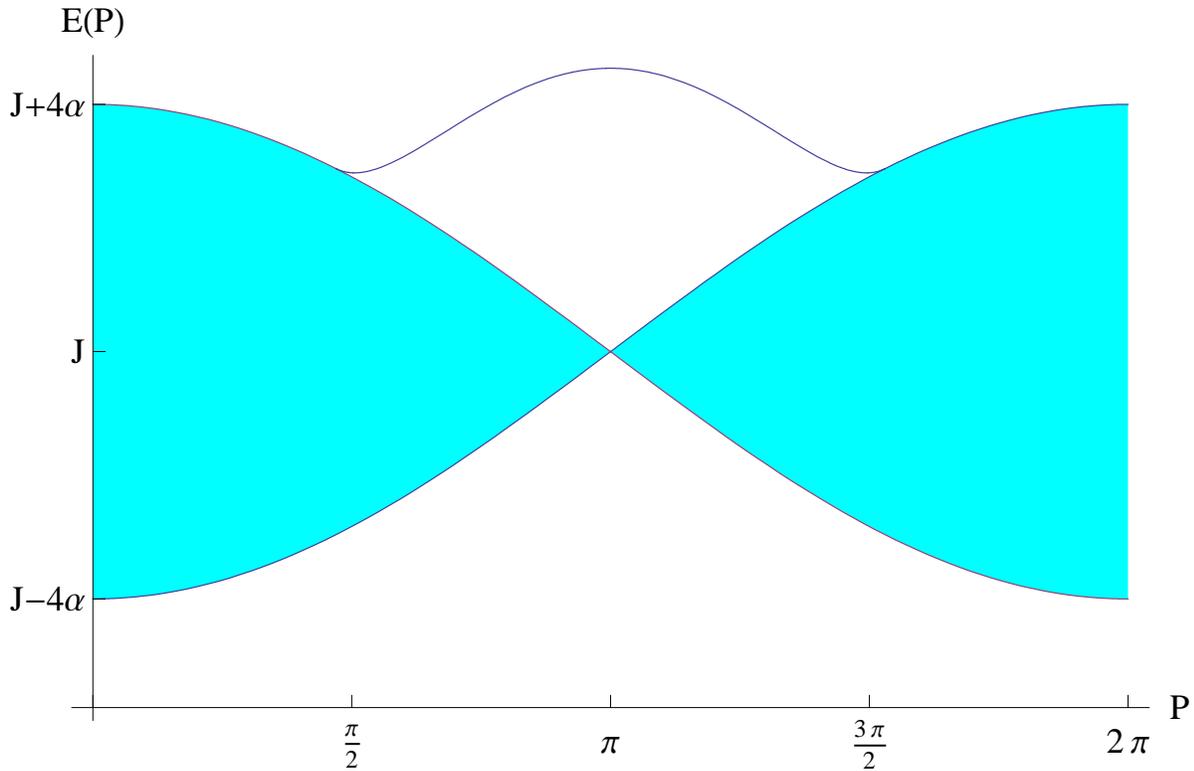}
\caption{\label{fig:sp0}  The energy spectrum of (\ref{HP}) at $h_z=0$, according to (\ref{cont}), (\ref{kin0}). 
The filled area corresponds to the continuous spectrum.  }
\end{figure}
At the edges of this region, the gap between the continuous spectrum and the kinetic mode vanishes.
The resulting energy spectrum at $h_z=0$ is shown in Figure \ref{fig:sp0}.

Returning to the original eigenvaule problem (\ref{HP}) with $h_z>0$, we rewrite it as follows
\begin{eqnarray}
\fl \left(-\lambda+\mu \,l+\frac{a\, \delta_{l,1} }{2}\right) \psi(l,P) 
 -\frac{\psi(l+1,P)+(1-\delta_{l,1})\psi(l-1,P)}{2}=0,
 \label{HP2}
\end{eqnarray}
where $l=1,2,\ldots$, the eigenfunction $\psi(l,P)$ vanishes at $l\to+\infty$, and 
\begin{eqnarray} \label{rsc}
\lambda= \frac{E(P)-J}{4\alpha \cos(P/2)} ,\quad a= \frac{\beta'-2\beta\cos P}{2\alpha \cos(P/2)},
\quad \mu=\frac{h_z}{4\alpha \cos(P/2)}.
\end{eqnarray}

It is possible to  extend equation (\ref{HP2})  to all integer 
$l\in {\mathbb Z}$, 
not necessary positive. 
Let us continue $\psi(P,l)$ skew-symmetrically to negative $l$ denoting
\begin{equation}
\Psi(l,P)=\cases{ \psi(l,P), & for\;$l=1,2,\ldots$ , \\0,& for\;$l=0,$\\
- \psi(-l,P,),& for\;$l=-1,-2,\ldots$  }
\end{equation}
One can easily check then, that the odd (in $l$) function $\Psi(l,P)$ solves equation
\begin{equation}
\left(-\lambda+\mu \,|l|+\frac{a\, \delta_{|l|,1} }{2}\right) \Psi(l,P) 
 -\frac{\Psi(l+1,P)+\Psi(l-1,P)}{2}=0,
 \label{HPsym}
\end{equation}
for $l=0,\pm1,\pm2,\ldots$, iff  the function $\psi(P,l)$ is the solution of the equation (\ref{HP2}) for $l=1,2,\ldots$

After the Fourier transform, equation (\ref{HPsym}) takes the form 
\begin{eqnarray} \label{Ht}
&&[\epsilon(z)-\lambda]\phi(z)+ \frac{a}{2}\Psi(1) \left(z-\frac{1}{z}\right) =-\mu z \oint_{S_1} \frac{\rmd z'}{ \pi \rmi}
\frac{\phi(z')}{(z'-z)^2}, \\ 
&&{\rm with }\;\; \label{eps}
\epsilon(z)=-\frac{1}{2} \left(z+\frac{1}{z}\right) ,\\
&&\Psi(1)=\oint_{S_1} \frac{\rmd z}{ 4\pi \rmi z}\phi_n(z)\left(\frac{1}{z}-z\right).
\end{eqnarray}
Here $S_1$ denotes the unit circle in the complex plane, the both $z$ and $z'$ variables lie in this circle, $|z|=|z'|=1$.
Integration in (\ref{Ht}) is taken in the counter-clockwise direction and understood in the sense of the 
Cauchy principal value.
The function $\phi(z)$ is defined as
\begin{equation} 
\phi(z) \equiv\sum_{l=-\infty}^\infty z^l\, \Psi(l,P)=\sum_{l=1}^\infty (z-z^{-1})\, \psi(l,P),
\end{equation}
and satisfies the symmetry property $\phi(1/z)=-\phi(z)$. We have dropped the explicit indication of the 
$P$-dependence in $ \phi(z)  $ and $\Psi(1)$, the full notation for these quantities should be $ \phi(z,P)  $ and $\Psi(1,P)$.

At $a=0$, problem (\ref{Ht}) reduces to the First Toy Model, solved in Subsection 6.1 of \cite{Rut08}. 
At $a\ne0$, problem (\ref{Ht}) can be solved by the same method. 
The result for the eigenvalues $\lambda_n$ reads as
\begin{equation}
\lambda_n=-\mu \,\nu_n ,
\end{equation}
where $\nu_n$ are the solutions of the equation
\begin{equation} \label{disp}
J_{\nu_n}(1/\mu) +a \,J_{\nu_n+1}(1/\mu)=0, 
\end{equation}
with $J_{\nu}(x)$ being the Bessel function of order $\nu$. In \ref{EigPr}, we describe
an alternative proof of this result. 

Accordingly, the solution of the  eigenvalue problem (\ref{ei}) reads as
\begin{equation}
E_n(P) =J-h_z\,\nu_n, \label{sp}
\end{equation}
and $\nu_n$  are the solutions of equation
\begin{equation} 
\fl 2\alpha\cos(P/2) J_{\nu_n}[ 4 h_z^{-1}\alpha \cos(P/2)] + ( \beta'-2\beta \cos P ) 
J_{\nu_n+1}[ 4 h_z^{-1}\alpha \cos(P/2)]=0.  \label{Bes}
\end{equation}

\begin{figure}[htb]
\centering
\includegraphics[width=\linewidth, angle=00]{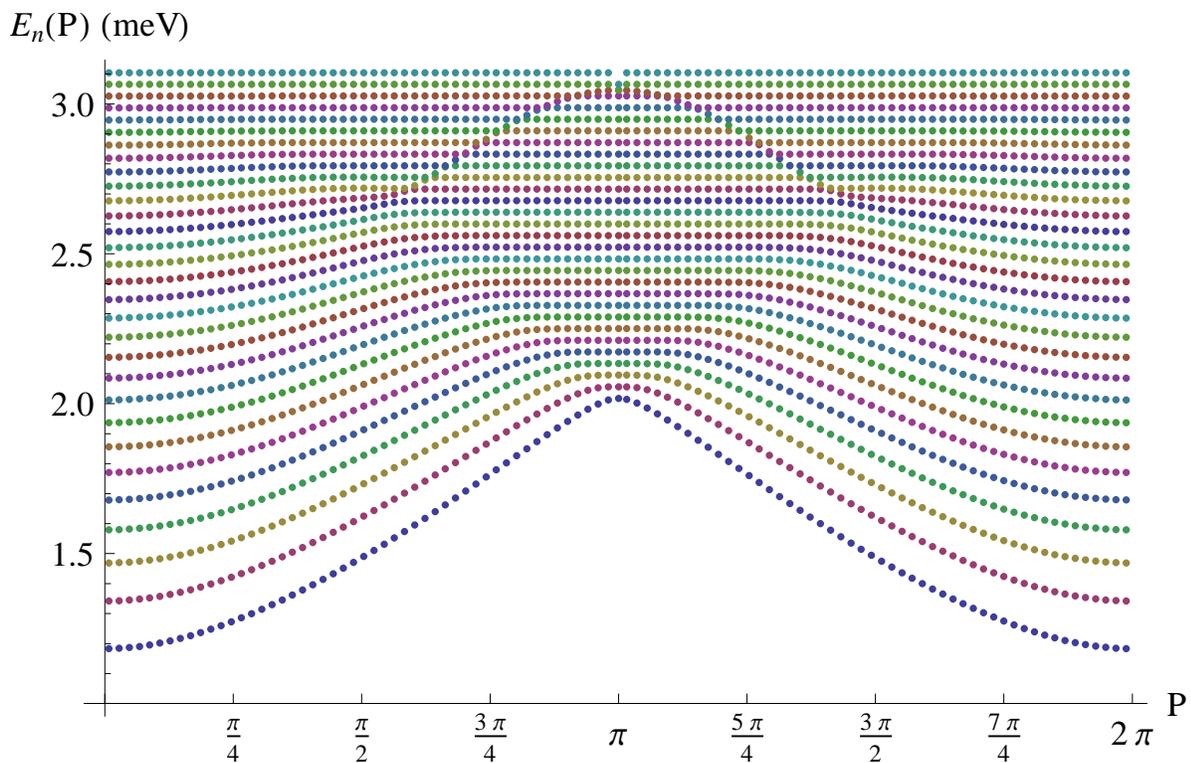}
\caption{\label{spe} Energy spectra of 30 lightest modes calculated from (\ref{sp}), (\ref{Bes}), (\ref{par}).}
\end{figure}

Figure \ref{spe} shows  30 lowest modes, calculated from (\ref{sp}), (\ref{Bes})
with the Hamiltonian parameter values 
\begin{equation} \label{par}
\fl J=1.94 \,\,{\rm meV}, \quad  \alpha=0.12 J,\quad \beta=0.17 \,J,\quad \beta'=0.21 \,J,\quad h_z=0.02 J,
\end{equation}
chosen by Coldea \etal \cite{suppC10}
to give the best fit of the experimental results.
\section{Weak-$h_z$ expansions \label{Sec3}}
If the longitudinal magnetic field is weak $h_z\to+0$, the long range 
confining potential $h_z l$ between the kinks
becomes small. Corresponding asymptotic expansions for the spectra $E_n(P)$ of their bound sates
can be extracted from the obtained exact results, either by means of  appropriate 
asymptotic formulas for the Bessel functions, or by direct calculation of
the integral in the left-hand side of equation [see equation (\ref{cstr})]
\begin{equation} \label{cst}
\int_{C} \rmd z'\,\left(\frac{1}{z'}+a\right) \exp\left\{\frac{\rmi}{\mu}[ {\mathcal F}(z') ]\right\} =0,
\end{equation} 
by the saddle point method at $\mu\to+0$. Here $ {\mathcal F}(z) = -\lambda \log z-\frac{1}{2}(z-z^{-1})$,
and contour $C$ is shown in  Fig. \ref{paths} in \ref{EigPr}. In this limit, 
one can distinguish several  asymptotical regimes for different $n$ and $P$, 
depending on the location of the saddle points 
\begin{equation}
z_{1,2}'=-\lambda_n\pm\sqrt{\lambda_n^2-1} \label{sad}
\end{equation}
in the $z'$-plane in integral (\ref{cst}). 
Since these calculations are very similar to those described in \cite{Rut08}, we present here only the 
final results.
\subsection{Low energy expansion}
If $n\sim 1$ and $P$ is well below the Brillouin zone boundary, $\lambda$ becomes close to $-1$,  
and the saddle points (\ref{sad})  merge at $z_{1,2}'=1$.   
In this case, one can use the 
low energy expansion in fractional powers of $h_z$:
\begin{eqnarray} 
E_n(P)=J-4 \alpha \cos(P/2) +[2\alpha \cos(P/2)]^{1/3} h_z^{2/3} z_n+  \label{low}\\
 \frac{h_z\, (\beta'-2\beta \cos P) }{ \beta'-2\beta \cos P +2\alpha \cos(P/2)}
-\frac{h_z^{4/3}z_n^2}{60\, [2 \alpha \cos(P/2)]^{1/3}}+O(h_z^{5/3}), \nonumber
\end{eqnarray}
with $(-z_n)$ being the zeros of the Airy function, ${\rm Ai}(-z_n)=0$.

The analogous expansion for $\lambda_n$ reads as
\begin{equation} 
\lambda_n=-1 + \frac {\mu^{2/3} z_n} {2^{1/3}} + \frac {\mu\, a} {1 + 
    a} - \frac {\mu^{4/3} z_n^2} {30\, 2^{2/3}}+O(\mu^{5/3}).
\end{equation}

The two leading term in expansion (\ref{low}) can be written in the form
\begin{equation}  
E_n(P)=\varepsilon(0,P) +\left[\frac{\partial^2}{\partial k^2}\Bigg|_{k=0} 
\frac{ \varepsilon(k,P) }{2}\right]^{1/3} h_z^{2/3} z_n+  \label{low1} 
O(h_z), \nonumber
\end{equation}
where $\varepsilon(k,P)$ is the continuous two-particle spectrum at $h_z=0$ given by (\ref{cont}).
This formula being in agreement with equation (87) of ref. \cite{Rut08}, is to a large extent 'model independent'. 
That is, its applicability does not depend on the explicit form of the two-particle spectrum $\varepsilon(k,P)$, 
provided that the factor in the square brackets in (\ref{low1}) is positive.
Note, that the second term in (\ref{low1}) explicitly depends on the bound-state momentum $P$, in contrast to the
finite momentum formula proposed by Coldea \etal to generalize equation (3) of ref. \cite{Coldea10},
see the in line equation for $m_j(k)$ in Page 8 of \cite{suppC10}. 
\subsection{Semiclassical expansions}
If $\lambda_n$ is deep inside the interval $(-1,1)$, 
the saddle points (\ref{sad}) shift from the real axis into the complex circle $|z'|=1$, and 
become two well separated complex conjugate numbers. 
Then $n\gg1$ at $h_z\to+0$, and one can easily obtain the semiclassical expansion for $E_n(P)$ from 
the saddle-point  asymptotics of the integral (\ref{cst}). 
To the leading order in $\mu \sim h_z$, we get
\begin{eqnarray} \label{sem}
E_n(P)=J-4\lambda_n \,\alpha\, \cos(P/2),\\
\lambda_n=-\cos\theta_n,\nonumber\\
\sin\theta_n-\theta_n \cos\theta_n=\mu\,\pi\,(n-1/4)+\mu\, \arg(1+a\, e^{i\theta_n})+O(\mu^2),\nonumber
\end{eqnarray}
where parameters $a$ and $\mu$ are given by (\ref{rsc}).

On the other hand, if $\lambda_n$ is well above  $1$, integral in (\ref{cst}) being determined by 
the saddle point $-\lambda_n+\sqrt{\lambda_n^2-1}$, lying in the real interval $(-1,0)$. 
However, there are still two saddle point
contributions to the integral in (\ref{cst}) coming from the upper and lower edges of the contour 
$C$, see Fig. \ref{paths}.  Since these contribution differ only by the phase 
factors $\exp(\mp i\pi \lambda_n/\mu)$, 
equation (\ref{cst}) can be written at $\mu\to+0$ with exponential 
accuracy as 
\begin{equation} \label{zer}
\sin\left(\frac{\pi\lambda_n }{\mu}\right)\,\int_{-1}^0 
\rmd z\,(a+z^{-1})\,\exp\left[\frac{-\lambda_n \log|z|-(z-z^{-1})/2}{\mu}\right]\approx 0.
\end{equation} 
Equating to zero the first factor in (\ref{zer}) leads the equidistant Zeeman ladder
\begin{eqnarray} \label{zee}
\lambda_n=\mu\,(n+1),\\
E_n(P)=J+h_z (n+1), \label{zee1}
\end{eqnarray}
with $n=1,2,3,\ldots$ Corresponding bound states can be considered semiclassically
as two well separated localized kinks moving back and forth without mutual collisions \cite{Rut08}.

Putting to zero the integral in (\ref{zer}) gives the energy of the well localized kinetic mode (\ref{loc})
modified by the longitudinal magnetic field $h_z$:
\begin{eqnarray} \nonumber
&&\lambda_{kin}=\frac{1+a^2}{2 a}+\frac{a^2\,\mu}{a^2-1}+O(\mu^2),\\
&& E_{kin}(P) =J+(\beta'-2\beta \cos P) +\frac{ [2\alpha \cos(P/2)]^2 }{\beta'-2\beta \cos P }+\label{kin}\\
&&h_z \frac{(\beta'-2\beta \cos P)^2}{ (\beta'-2\beta \cos P)^2 -[2\alpha \cos(P/2)]^2}  \nonumber
 +O(h_z^2).
\end{eqnarray}

\begin{figure}[htb]
\centering
\includegraphics[width=\linewidth, angle=00]{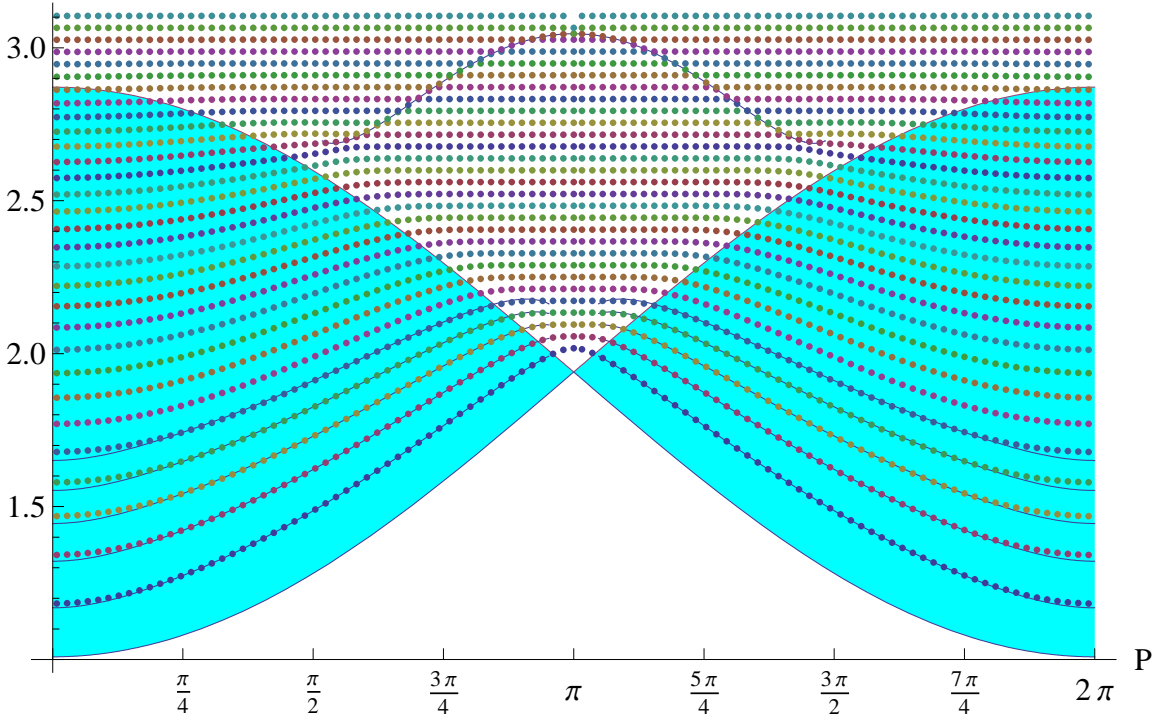}
\caption{\label{as} The same spectra as in Fig. \ref{spe} with added asymptotical curves explained in the text.}
\end{figure}

Fig. \ref{as} shows the same spectra as in Fig. \ref{spe}, together with six asymptotical curves. 
The filled area displays the region of the continuous spectrum at $h_z=0$.
Five curves in the bottom  display the spectra of five lightest modes calculated from 
the low-energy expansion (\ref{low}). This expansion does not hold outside the filled region. 
The curve crossing the Zeeman ladder in the top 
of Fig. \ref{as}
represents the kinetic bound-state mode determined from (\ref{kin}). It is clear, however, that this  
are in fact the avoided crossings with exponentially narrow gaps. 

Note, that all above asymptotical formulas cannot be used in the crossover region near the 
upper bound of the filled area.
\section{Dynamical correlation function \label{Sec4}}
Besides the magnetic excitation energy spectra, the inelastic neutron scattering allows one to measure 
the dynamical correlation function \cite{Coldea10,suppC10}
\begin{equation} \label{dcf}
S^{xx}(P,\omega) = S^{yy}(P,\omega)=\sum_{n=1}^\infty |\langle \Phi_n(P)|S_0^x|0\rangle|^2\,\delta[\omega-E_n(P)],
\end{equation}
where $ |0\rangle = | \ldots \uparrow \uparrow\uparrow\ldots \rangle$ is the ferromagnetic ground state,
   $S_j^x=\sigma_j^x/2$ is the $x$-spin operator at the site $j$, and the sum extends over all 
eigenstates of the Hamiltonian with the momentum $P$. Since the operator $S_j^x$ inverts just one spin at the 
site $j$ in the chain, it maps the ferromagnetic vacuum into the two-kink state, 
\begin{equation*}
S_j^x|0\rangle  \equiv S_j^x   |\ldots \uparrow \uparrow \stackrel{j}{\uparrow}\uparrow \uparrow\ldots\rangle = 
 \frac{1}{2}\,
|\ldots\uparrow  \uparrow \stackrel{j}{\downarrow}\uparrow\uparrow \ldots\rangle\equiv
 \frac{1}{2}\,|j,1\rangle.
\end{equation*}
Accordingly, the dynamical correlation function (\ref{dcf}) for model (\ref{H}) takes the form 
\begin{eqnarray}
S^{xx}(P,\omega) =\frac{1}{4} \sum_{n=1}^\infty I_n(P)  \,\delta[\omega-E_n(P)],\\
\sum_{n=1}^{\infty} I_n(P) =1, \nonumber
\end{eqnarray}
where $I_n(P)$ is the  relative intensity of the $n$-th mode, 
\begin{equation}
I_n(P) =\frac{| \psi_n(l=1,P) |^2}{\sum_{l=1}^\infty | \psi_n(l,P) |^2 }. \label{intens}
\end{equation}
and $\psi_n(l,P)$ are the  eigenfunctions of  Hamiltonian (\ref{HP}).

Equations (\ref{dcf})-(\ref{intens}) are written in the assumption that Hamiltonian (\ref{H}) 
has only the discrete spectrum. 
These relations can be easily  modified in the case $h_z=0$, where the continuous spectrum exists:
\begin{eqnarray}
\fl S^{xx}(P,\omega) =\frac{1}{4} I_{kin}(P)  \,\delta[\omega-E_{kin}(P)]+
\frac{1}{2\pi} \int_{0}^\pi \rmd k \, I(k,P)  \,\delta[\omega-E(k,P)],\\
\fl I_{kin}(P)   +\frac{2}{\pi} \int_{0}^\pi \rmd k\, I(k,P) =1, \nonumber
\end{eqnarray}
where
\begin{eqnarray} \label{ikin}
I_{kin}(P) =\left\{
\begin{array}{ll}
 0, & {\rm if } \quad |a|\le 1 \\
1-a^{-2},  & {\rm if }\quad|a|> 1,
\end{array}
\right.\\
I(k,P) =\sin^2\gamma=\frac{\sin^2 k}{1+2 a \cos k+a^2}, \label{inc}
\end{eqnarray}
and parameters $a$ and $\gamma$  are given by (\ref{rsc}) and (\ref{gamma}), respectively.
\begin{figure}[htb]
\centering
\includegraphics[width=.7\linewidth, angle=00]{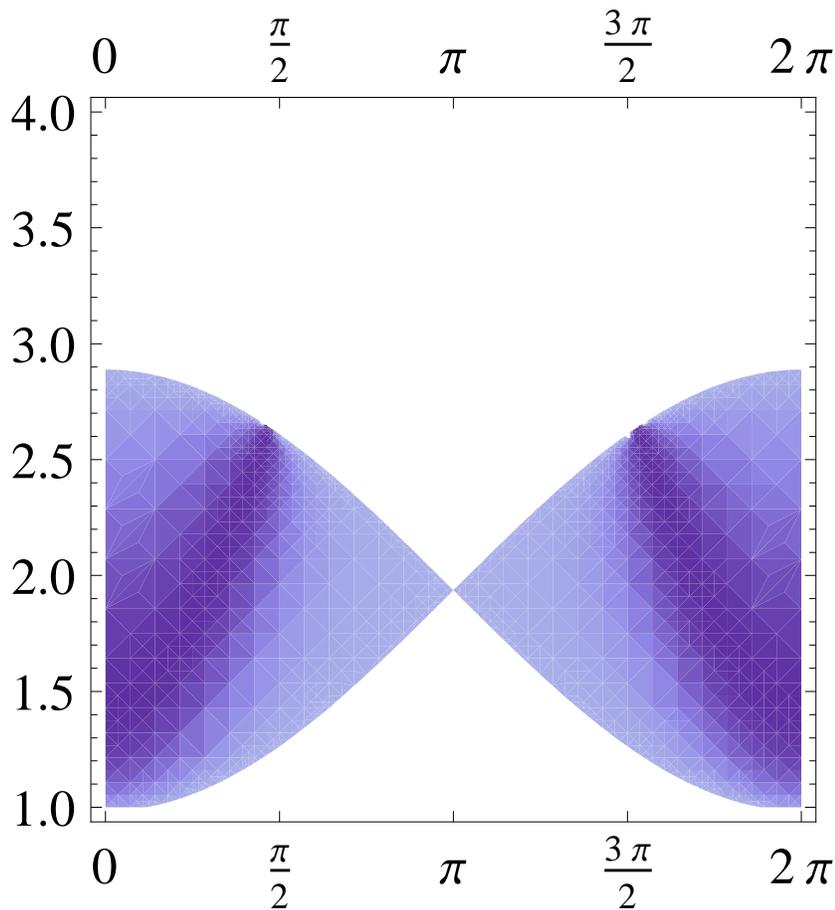}
\caption{\label{in0} Intensity $I(k,P)$ of the continuous modes at $h_z=0$ in the $(P,E)$-plane 
determined from (\ref{inc}), (\ref{gamma}). Darker regions correspond to larger intensity.}
\end{figure} 
Fig. \ref{in0} displays the density plot of the intensity $I(k,P)$ of the continuous modes at $h_z=0$ in the $(P,E)$-plane 
determined from (\ref{inc}), (\ref{gamma}) for the values of the rest parameter given by (\ref{par}).

Formulae (\ref{dcf})-(\ref{intens}) can be directly used at nonzero longitudinal magnetic field $h_z>0$, 
since the spectrum is discrete in this case. Using the results of \ref{Int}, the relative intensities 
of the discrete modes can be expressed in terms of the Bessel function
\begin{eqnarray} \label{ikinh}
I_{n}(P)  =2\mu
\left\{\frac{\partial}{\partial \nu}\left[\frac{J_\nu(1/\mu)}{J_{\nu+1}(1/\mu)}\right]\right\}^{-1} \bigg|_{\nu\to\nu_n},
\end{eqnarray}
where $\mu$ is given by (\ref{rsc}), and $\nu_n$ is the $n$-th solution of equation (\ref{disp}).

\begin{figure}[htb]
\centering
\includegraphics[width=\linewidth, angle=00]{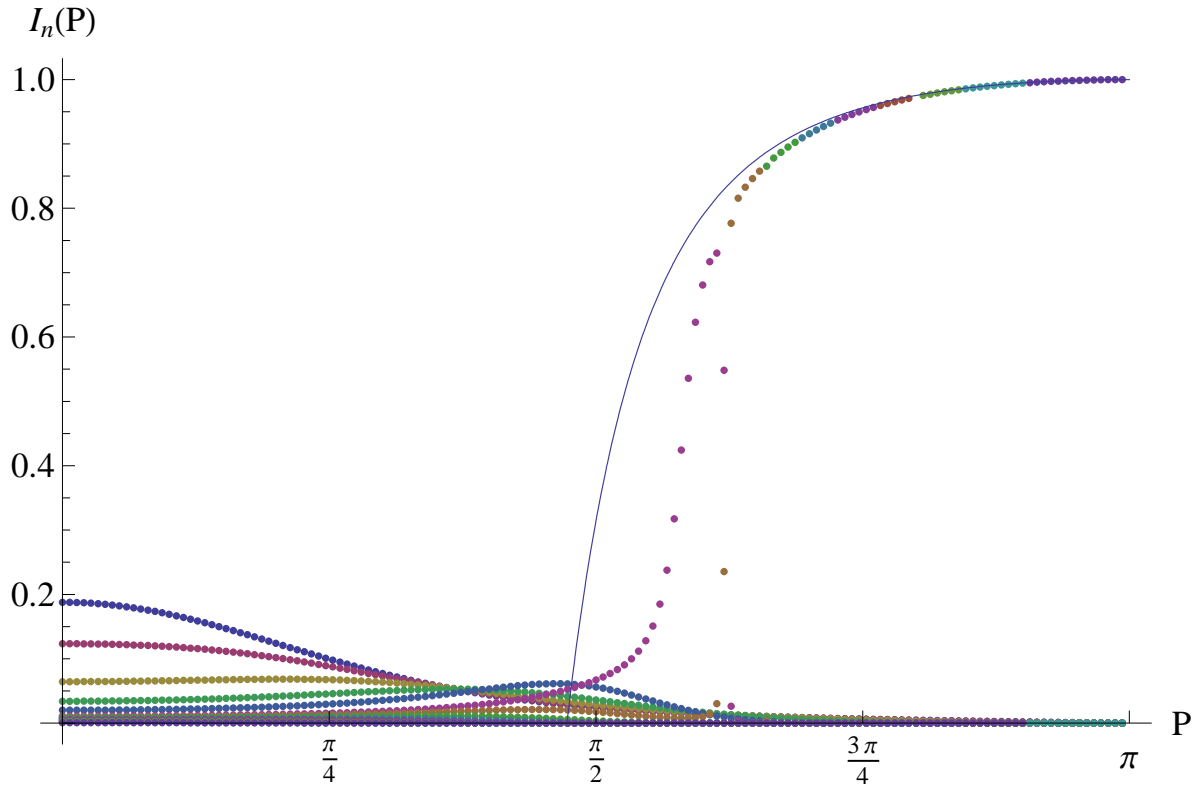}
\caption{\label{fig:inth} Intensities of the discrete modes $I_n(P)$ at a weak 
longitudinal magnetic field according to (\ref{ikinh}),
with the parameter values (\ref{par}). The solid curve represents the intensity
$I_{kin}(P)$ of the kinetic mode at $h_z=0$, according to (\ref{ikin}).  }
\end{figure}
\begin{figure}[htb]
\centering
\includegraphics[width=\linewidth, angle=00]{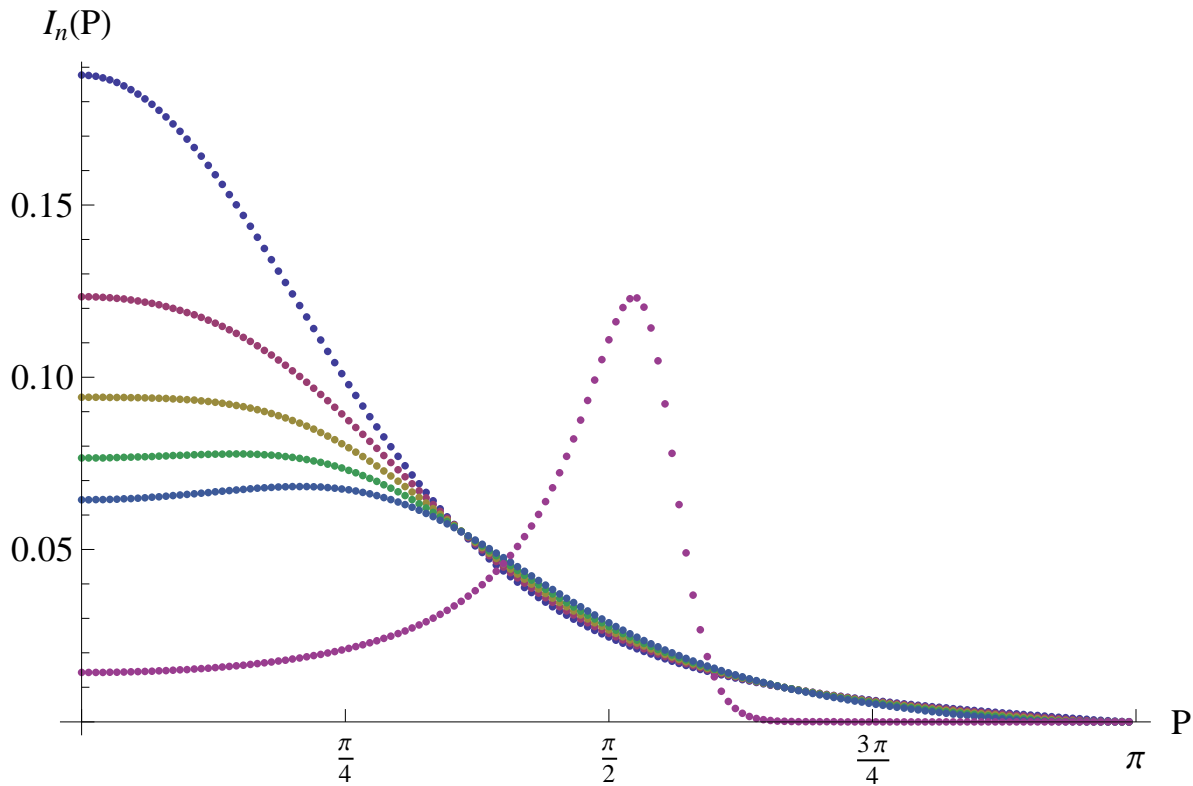}
\caption{\label{fig:inth5} Intensities $I_n(P)$ for modes with $n=1,2,3,4,5,18$ at 
nonzero $h$ according to (\ref{ikinh}), (\ref{rsc}), (\ref{par}).
Intensities $I_n(0)$ decrease with increase $n$. }
\end{figure}

Figures \ref{fig:inth} and  \ref{fig:inth5} show  intensities $I_n(P)$ of  discrete 
modes calculated from (\ref{ikinh}), (\ref{rsc}) with the parameter values
(\ref{par}). 
Intensities $I_n(0)$ decrease with increasing $n$. Intensities of 30 lightest modes are plotted versus
the momentum $P$
in Fig. \ref{fig:inth}, and those for $n=1,2,3,4,5,18$ are shown in Fig. \ref{fig:inth5}. 
 The intensity of the 
kinetic mode $I_{kin}(P)$ at $h_z=0$ is shown in Fig. \ref{fig:inth} by the solid curve. 

\begin{figure}[htb]
\centering
\includegraphics[width=\linewidth, angle=00]{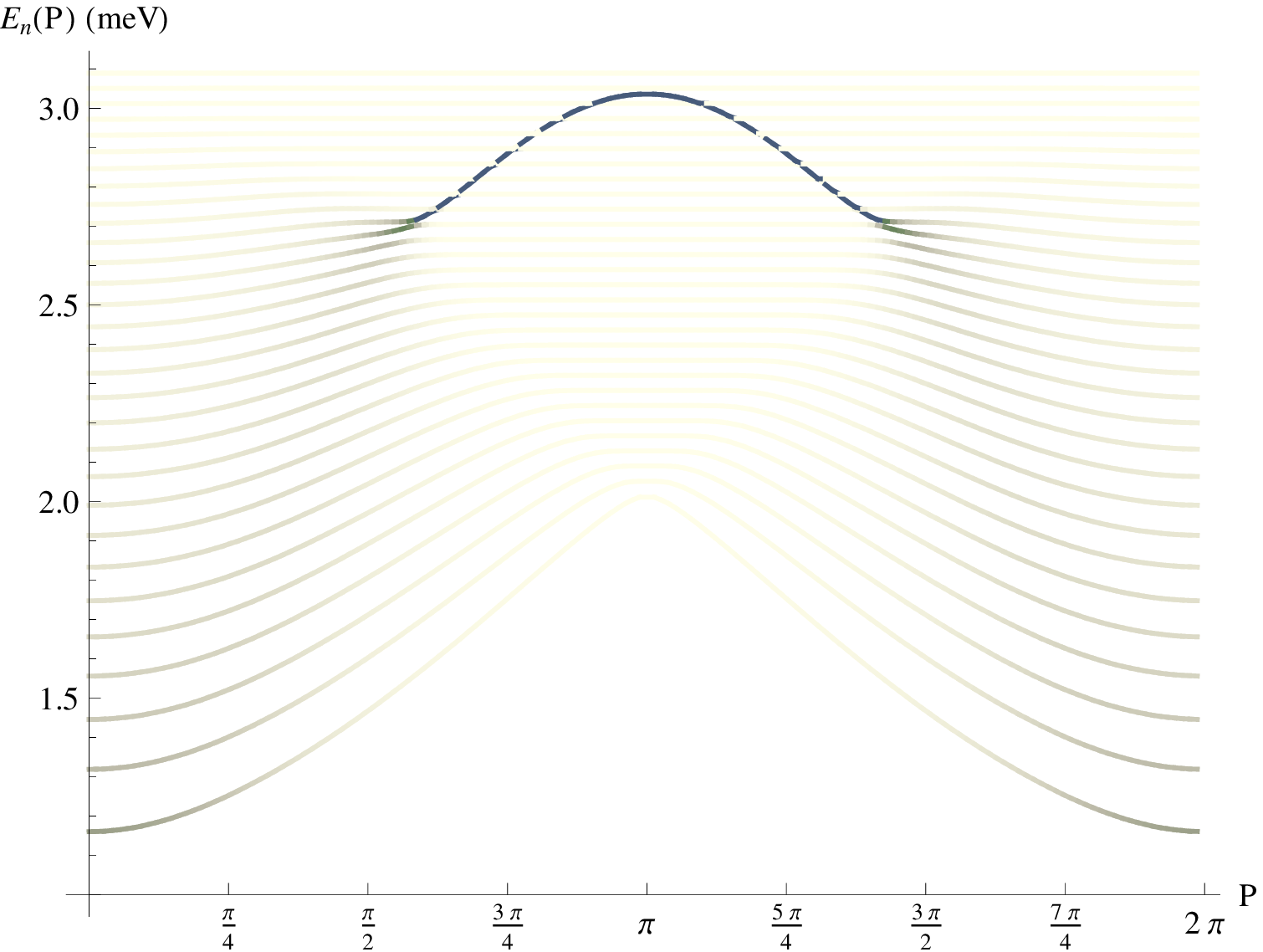}
\caption{\label{fig:IntAll} Energy spectra of 30 lowest modes according to (\ref{sp}), (\ref{Bes}), (\ref{par}).
Darkness of the curves characterizes the 
intensities (\ref{ikinh}) of the modes. }
\end{figure}

The obtained results are summarized in Fig. \ref{fig:IntAll}. It shows the same energy spectra 
as those in Fig. \ref{spe}. The dispersion curves were calculated by use of equations 
(\ref{sp})-(\ref{par}). 
The darkness of the dispersion curves in Fig. \ref{fig:IntAll} characterizes the 
relative intensities
(\ref{ikinh}) for the corresponding modes. This figure looks quite similar to
Fig. 3B of ref. \cite{Coldea10}, which displays the experimentally observed neutron scattering spectra in
$\mathrm{Co}\mathrm{Nb}_2\mathrm{O}_6$.
\section{Conclusions \label{Conc}} 
In this paper we obtain the exact solution of the phenomenological model, which was
proposed  by Coldea \etal \cite{suppC10} to describe the spectra of two-kink bound states in the entire 
Brillouin zone observed in the quasi-1D Ising ferromagnet
$\mathrm{Co}\mathrm{Nb}_2\mathrm{O}_6$ in the inelastic neutron scattering experiments \cite{Coldea10}.
The model Hamiltonian acts in the space of two-kink configurations of the spin chain, and parametrizes hoppings of kinks
along the chain, their short range interaction, and the linear long-range attracting potential leading to  confinement
of kink into pairs. We express the dispersion law $E_n(P)$ of two-kink bound states and relative neutron scattering 
intensities $I_n(P)$ by the magnetic modes in terms of the Bessel function. The preliminary analysis shows \cite{pcC10}, 
that the obtained spectra $E_n(P)$ plotted in Fig. \ref{spe} display a good 
quantitative agreement with the experimental data.  
It would be interesting to perform the detailed comparison of obtained exact solutions [both for energies $E_n(P)$ and
intensities $I_n(P)$]
with the neutron scattering data in $\mathrm{Co}\mathrm{Nb}_2\mathrm{O}_6$. 

Let us note, that the dispersionless modes forming the almost equidistant ladder  with energies (\ref{zee1})
and momenta near the zone boundary have very small intensities $I_n(P)$, as it is seen from Fig. \ref{fig:IntAll}. 
The reason is quite simple. It is well known, that the linear potential causes localization of a sole kink in 
the discrete spin chain, similarly to localization of an electron in the isolated conducting zone
by the uniform electric field \cite{zim72}. The dispersionless modes (\ref{zee1}) correspond to 
large enough clusters of down-spins bounded by two well separated localized kinks,
which can be hardly excited in the up-spin ground state by the neutrons  scattering. 
Perhaps, such modes could be effectively excited by application of the oscillating magnetic field having the 
resonant frequency $\omega= h_z/\hbar$.

\noindent
\ack 
I am thankful to R. Coldea for interesting correspondence. \newline
\noindent
This work 
is supported by  the Belarusian 
Republican Foundation for Fundamental Research.  
\appendix
\section{Solution of the eigenvalue problem \label{EigPr} }
In this Appendix we describe the exact solution of the eigenvalue problem (\ref{HP2}), which we rewrite as
\begin{equation}
\fl \left(-\lambda+\mu \,l+\frac{a\, \delta_{l,1} }{2}\right) \psi(l) 
 -\frac{\psi(l+1)+\psi(l-1)}{2}=0, \quad\quad l=1,2,\ldots,
 \label{HPA}
\end{equation}
with the Dirichlet boundary conditions
\begin{eqnarray}
\lim_{l\to+\infty}\psi(l)=0,\\
\psi(0)=0. \label{restr}
\end{eqnarray}
The following normalization condition for the eigenstate $\psi(l)$ will be chosen:
\begin{equation}
\psi(1)=-2. \label{norm}
\end{equation}

Let us skip for a while the boundary condition (\ref{restr}) and consider the generating function $g(z)$ for $\psi(l)$,
\begin{equation} \label{gen}
g(z)=\sum_{l=1}^\infty\psi(l)\,z^l.
\end{equation}
This function should be analytical inside the circle $|z|<1$, 
subject to the boundary condition 
\begin{equation} \label{bc}
g(0)=0, 
\end{equation}
and satisfy the differential equation 
\begin{equation}\label{dif}
[\epsilon(z)-\lambda]g(z)+\mu\, z \frac{dg(z)}{dz} =V(z),
\end{equation}
following directly from (\ref{HPA}), (\ref{norm}), (\ref{gen}). 
Here $\epsilon(z)$ is given by (\ref{eps}), and 
\begin{equation} \label{V}
V(z)=1+ z\left[a+\frac{\psi(0)}{2}\right].
\end{equation}
The solution of equation (\ref{dif}) satisfying (\ref{bc}) reads as
\begin{equation} \label{sol}
g(z)=\int_0^z \frac{\rmd z'}{\mu z'}\,V(z') \exp\left\{\frac{\rmi}{\mu}[ {\mathcal F}(z') - {\mathcal F}(z) ]\right\},
\end{equation}
where
\begin{equation}
\rmi \, {\mathcal F}(z) = -\lambda \log z-\frac{1}{2}(z-z^{-1}),
\end{equation}
and the branch of the logarithm is fixed by the condition ${\mathcal F}(1)=0$.
\begin{figure}[htb]
\centering
\includegraphics[width=.7\linewidth, angle=00]{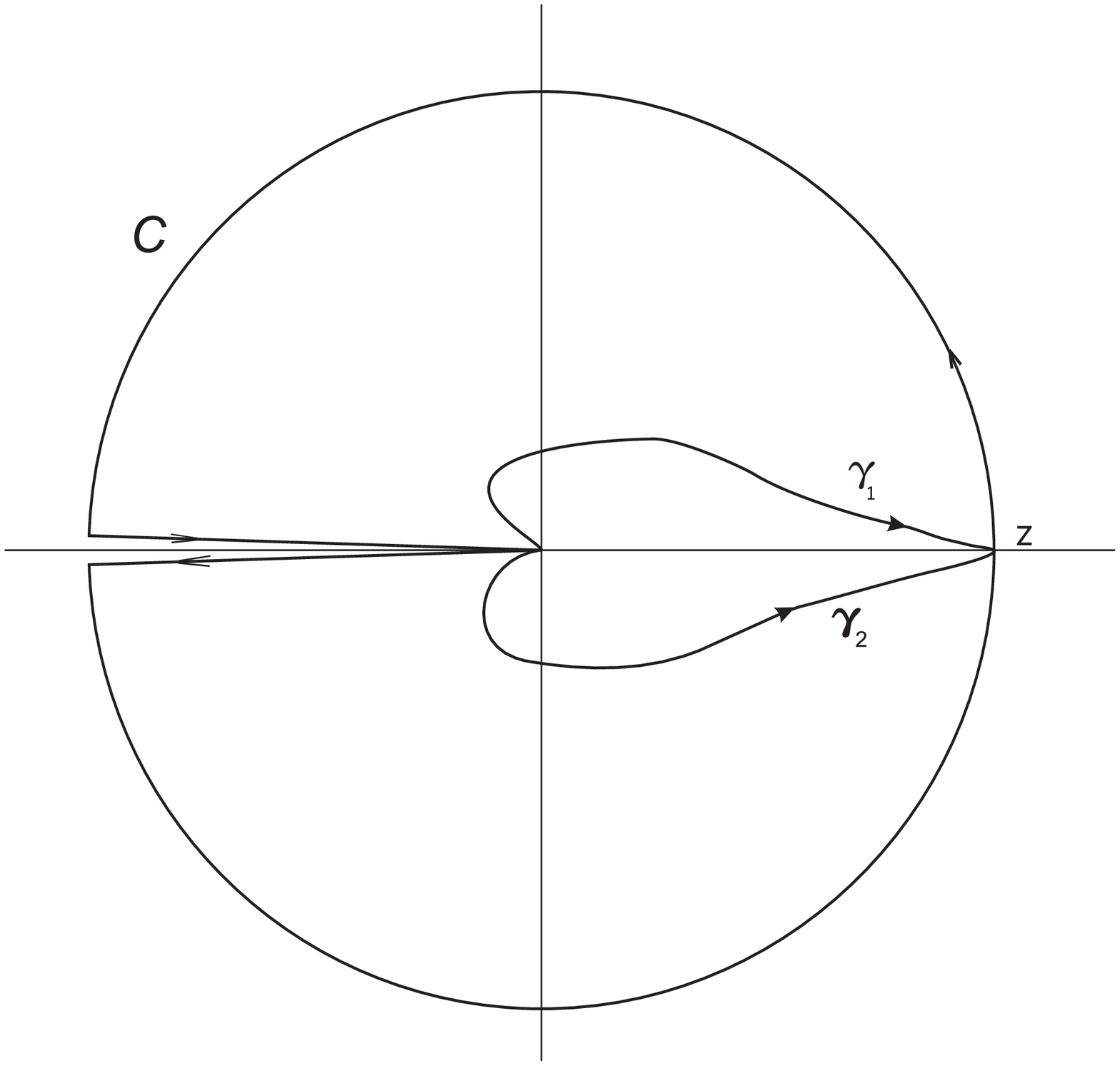}
\caption{\label{paths} Integration paths in the $z'$-plane: arcs $\gamma_1$ and $\gamma_2$ connecting the points $z'=0$ 
and $z'=z$ are the 
allowed integration paths in  (\ref{sol}); loop $C$ is the integration path in (\ref{cstr}). }
\end{figure}

At positive $\mu$, the integration path in the $z'$-plane 
in (\ref{sol}) should approach the origin $z'=0$ from the left half-plane ${\mathrm Re}\, z'<0$
to provide convergence of the integral. Figure \ref{paths} shows two 
allowed integration paths $\gamma_1$ and $\gamma_2$ in the 
right-hand side of (\ref{sol}) for the case $z>0$. Integration along each of them should give the same function $g(z)$.
This requirement  leads to the following constraint (cf. equations (56), (57)  in ref. \cite{Rut08})
\begin{equation} \label{cstr}
\int_{C} \frac{\rmd z'}{ z'}\,V(z') \exp\left\{\frac{\rmi}{\mu}[ {\mathcal F}(z')]\right\}=0,
\end{equation} 
where the integration path $C=\gamma_2-\gamma_1$ is shown in Figure \ref{paths}. 
Substitution of (\ref{V}) into (\ref{cstr}) gives the constant $\psi(0)$:
\begin{equation} \label{psi0}
\psi(0)=-2 \left[ a+\frac{J_\nu(\mu^{-1})}{J_{\nu+1}(\mu^{-1})} \right],
\end{equation}
where $\nu=-\lambda/\mu$, and $J_\nu(x)$ is the Bessel function. We have taken into account the 
integral representation of the Bessel function
\[
 J_\alpha(x)=\int_C \frac{\rmd z}{2\pi \rmi\,z} \,z^{\alpha}\exp[x(z^{-1}-z)/2],
\]
which reduces to the well known form (see formula (5) in page 15 in \cite{Bat}) 
after the change of the integration variable $z=1/u$.

Application of  the Dirichlet boundary condition (\ref{restr}) leads then to 
the equation 
\begin{equation} \label{cstr1}
J_{\nu} (\mu^{-1})+ a\,J_{{\nu}+1} (\mu^{-1})  =0,
\end{equation}
which solutions $\nu_n$ determine the spectrum $\lambda_n=-\mu\, \nu_n$ of the problem (\ref{HPA})-(\ref{restr}).

\begin{figure}[htb]
\centering
\includegraphics[width=.7\linewidth, angle=00]{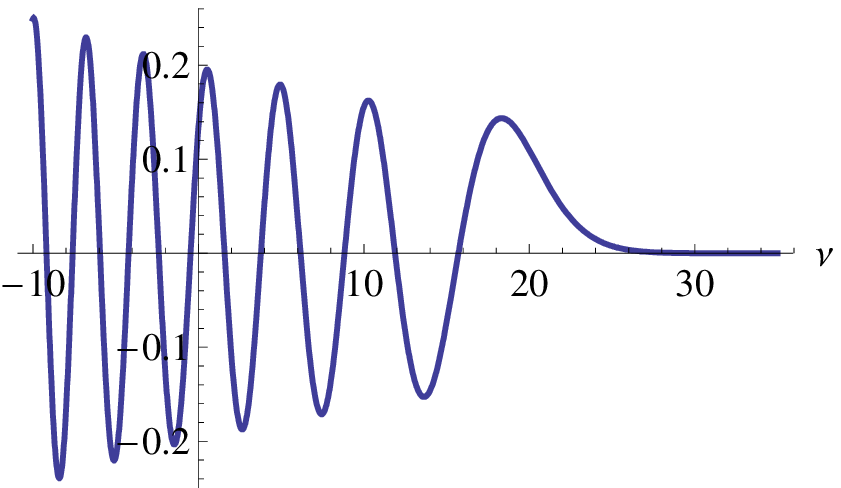}
\caption{\label{bess} Plot of the function 
$J_{\nu} (\mu^{-1})+ a\,J_{{\nu}+1} (\mu^{-1}) $ versus $\nu$ at $a=-0.5$, $\mu=0.05$.}
\end{figure}

Fig. \ref{bess} shows the plot of the left-hand side of equation (\ref{cstr1}) versus $\nu$ at $a=-0.5$, $\mu=0.05$.
\section{Relative intensities of discrete modes at $\mu>0$ \label{Int}}
In this Appendix we  prove the following \newline
{\underline {Statement.}}\newline 
Let the set of real numbers $\{\psi(l,\lambda)\}_{l=0}^\infty $ solves the linear problem 
\begin{equation}
\fl \left(-\lambda+\mu \,l+\frac{a\, \delta_{l,1} }{2}\right) \psi(l,\lambda) 
 -\frac{\psi(l+1,\lambda)+\psi(l-1,\lambda)}{2}=0, \quad\quad l=1,2,\ldots,
 \label{HPB}
\end{equation}
with the boundary condition
\begin{equation}
\lim_{l\to+\infty}\psi(l,\lambda)=0. \label{BC}
\end{equation} 
Coefficients $a,\lambda, \mu$ are supposed to be real, and $\mu>0$.\newline
Let us define the intensity $I(\lambda)$ corresponding to this solution as
\begin{equation} \label{def}
I(\lambda) =\frac{[ \psi(1,\lambda)]^2 }{\sum_{l=1}^\infty[ \psi(l,\lambda)]^2}.
\end{equation}
Then
\begin{equation} \label{int}
I(\lambda)=2 \mu \left[ \frac{\partial}{\partial \nu} 
\frac{ J_\nu(\mu^{-1}) }{J_{\nu+1}(\mu^{-1})} \right]^{-1} \Bigg|_{\nu=-\lambda/\mu}.
\end{equation}
{\underline {Proof.}}\newline
Since definition (\ref{def}) of $I(\lambda)$ does not depend on 
the normalization of the set $\{\psi(l,\lambda)\}_{l=0}^\infty$, we shall
fix the latter by the condition 
\begin{equation} \label{normB}
\psi(1,\lambda) =-2, 
\end{equation}
without loss of generality.
Consider  the solution $\{\psi(l,\lambda')\}_{l=0}^\infty$ of the problem (\ref{HPB}), 
(\ref{BC}), (\ref{normB}) in which $\lambda$ is replaced by 
$\lambda'$. In particular, instead of (\ref{HPB}) we get
\begin{equation}
\fl \left(-\lambda'+\mu \,l+\frac{a\, \delta_{l,1} }{2}\right) \psi(l,\lambda')
 -\frac{\psi(l+1,\lambda')+\psi'(l-1,\lambda')}{2}=0, \quad\quad l=1,2,\ldots
 \label{HPB2}
\end{equation}
Let us multiply equation (\ref{HPB}) by $\psi(l,\lambda')$,  and equation (\ref{HPB2}) by $\psi(l,\lambda)$, 
then subtract one equation from the another and sum the result over all natural $l$. As the result, we obtain
\begin{equation}
(\lambda'-\lambda)\sum_{l=1}^{+\infty} \psi(l,\lambda)\,\psi(l,\lambda') =\frac{1}{2}
\left[ \psi(1,\lambda')\psi(0,\lambda) -\psi(1,\lambda)\psi(0,\lambda')\right],
\end{equation}
or
\begin{equation} \label{sum}
\sum_{l=1}^{+\infty} \psi(l,\lambda)\,\psi(l,\lambda') = \frac{1}{2(\lambda'-\lambda)}
\left[ \psi(1,\lambda')\psi(0,\lambda) -\psi(1,\lambda)\psi(0,\lambda')\right].
\end{equation}
Taking into account the chosen normalization condition $ \psi(1,\lambda) =\psi(1,\lambda')=-2$,
 and proceeding in (\ref{sum}) to the limit $\lambda'\to\lambda$, we get
\begin{equation} \label{sum3}
\sum_{l=1}^{+\infty} [\psi(l,\lambda)]^2 =\frac{d \psi(0,\lambda)}{d\lambda}.
\end{equation}
Combining (\ref{sum3}) with (\ref{psi0}) yields
\begin{equation} \label{s2}
\sum_{l=1}^{+\infty} [\psi(l,\lambda)]^2=\frac{2}{\mu}\,\left[\frac{\partial}{\partial \nu} 
\frac{ J_\nu(\mu^{-1}) }{J_{\nu+1}(\mu^{-1})}
\right] 
\Bigg|_{\nu=-\lambda/\mu}.
\end{equation}
Substitution of (\ref{normB}) and (\ref{s2}) into (\ref{def}) leads finally to the result (\ref{int}).
\section*{References}
\bibliography{BankIsing}
\end{document}